\documentclass[twoside,american,conference]{IEEEtran}
\usepackage[T1]{fontenc}
\usepackage[latin9]{inputenc}
\usepackage{float}
\usepackage{bm}
\usepackage{amsmath}
\usepackage{amsthm}
\usepackage{amssymb}
\usepackage{graphicx}

\makeatletter

\newcommand{\noun}[1]{\textsc{#1}}
\providecommand{\tabularnewline}{\\}
\floatstyle{ruled}
\newfloat{algorithm}{tbp}{loa}
\providecommand{\algorithmname}{Algorithm}
\floatname{algorithm}{\protect\algorithmname}

\theoremstyle{plain}
\newtheorem{thm}{\protect\theoremname}

\usepackage{graphicx, amsmath}
\usepackage{subfigure}
\usepackage{cite}
\usepackage{algorithmic}

\renewcommand{\fnum@figure}{Fig.~\thefigure}

\makeatother

\usepackage{babel}
\providecommand{\theoremname}{Theorem}

\begin{document}

\title{Energy Efficient Precoding C-RAN Downlink with Compression at Fronthaul}

\author{\IEEEauthorblockN{Kien-Giang Nguyen$^{\ast}$, Quang-Doanh Vu$^{\ast}$, Markku Juntti$^{\ast}$,
and Le-Nam Tran$^{\dagger}$} \IEEEauthorblockA{$^{\ast}$ Centre for Wireless Communications, University of Oulu,
P.O.Box 4500, FI-90014 University of Oulu, Finland\\
Email: \{kien.nguyen, quang.vu, markku.juntti\}@oulu.fi\\
$^{\dagger}$ Department of Electronic Engineering, Maynooth University,
Ireland; Email: lenam.tran@nuim.ie}}
\maketitle
\begin{abstract}
This paper considers a downlink transmission of cloud radio access
network (C-RAN) in which precoded baseband signals at a common baseband
unit are compressed before being forwarded to radio units (RUs) through
limited fronthaul capacity links. We investigate the joint design
of precoding, multivariate compression and RU-user selection which
maximizes the energy efficiency of downlink C-RAN networks. The considered
problem is inherently a rank-constrained mixed Boolean nonconvex program
for which a globally optimal solution is difficult and computationally
expensive to find. In order to derive practically appealing solutions,
we invoke some useful relaxation and transformation techniques to
arrive at a more tractable (but still nonconvex) continuous program.
To solve the relaxation problem, we propose an iterative procedure
based on DC algorithms which is provably convergent. Numerical results
demonstrate the superior of the proposed solution in terms of achievable
energy efficiency compared to existing schemes.
\end{abstract}

\section{Introduction }

The evolution of wireless communication techniques towards the foreseen
fifth generation (5G) wireless networks envisions a dramatic growth
of wireless devices, applications and demand on wireless data traffic
\cite{1000xdata,ericsson2011more}. Accordingly, s\emph{pectral efficiency}
(SE) will certainly play a major role in future cellular. As well
concluded in pioneer research, multicell cooperation or cooperative
multipoint processing (CoMP) with joint base station (BS) processing
and transmission  is a promising enabling technique to tackle the
`spectrum crunch' problem \cite{Multicell-MIMO-Cooperative-Networks,HowdesignCOMP}.
However, the transmission with large numbers of antenna elements
consumes remarkable amount of processing power or energy.  Thus \emph{energy-efficiency}
(EE) has appeared as another important design objective. Recently,
energy-efficient techniques and architectures for cooperative transmission
have been intensively investigated \cite{GreenWirelessComm}.

Among them, cloud radio access network (C-RAN) is appearing as a revolutionary
architectural solution to the problem of enhanced SE and EE requirements
for cellular networks \cite{GreenWirelessComm,mobile2011c}. In CRANs,
baseband (BB) signal processing components are no longer deployed
at base stations, but installed at a common BB unit (BBU), which is
now responsible for the encoding/decoding and other computational
tasks on transmitted signals. Thus CRANs can take full advantage of
the cooperative principle to boost the achievable SE. In addition,
conventional BSs are also replaced by low-cost low-power ones which
are equipped with only radio frequency modules, thereby reducing the
power cost for management and operating the BSs. Such a BS is often
called as\emph{ radio unit} (RU). Nevertheless, to perform the transmission/reception,
each RU has to receive/forward BB signals from/to the BBU through
\emph{fronthaul} links, which can be wired or wireless. In either
case, they are capacity limited and thus allow only a limited amount
of BB information to be transferred per a time unit. This limitation
explicitly restricts the performance of C-RANs and becomes a factor
that needs to be taken into consideration in the system design \cite{Fronthaul-constraintCRAN}.

This paper is focused on the downlink transmission of C-RAN such that
the BB signals are precoded at the BBU, which are then compressed
before being forwarded to RUs through fronthaul links. We aim at studying
a joint design of precoding, multivariate compression and RU-user
selection that maximizes the EE of the CRAN downlink under a limited
power budget and finite-capacity fronthaul links. More specifically,
joint precoding and multivariate compression design are adopted to
improve the system throughput \cite{JointprecodingMultivariate,FronthaulCompressionCRAN-Mag},
while a proper RU-user selection scheme potentially reduces the power
and fronthaul expenditure of the network. The considered problem is
cast as a rank-constrained mix-Boolean nonconvex program, which belongs
to a class of NP-hard problems, and thus a globally optimal solution
is hard to find. Therefore, we propose a low-complexity method that
solves the problem locally, which is a classical goal for such an
NP hard problem. To this purpose, we first drop the rank constraint
and lift the Boolean variables into the continuous domain. Then by
using novel transformations, we show that the relaxed problem admits
a difference of convex (DC) function structure, which motivates the
application of DC algorithms \cite{MarksWright:78:AGenInnerApprox,BBT:10:JGO,le2014dc}
to achieve suboptimal solutions. Particularly, the problem is convexified
into a semidefinite program (SDP) at each iteration of the proposed
algorithm using the principle of the DC programming. This produces
a sequence of iterates which provably converges to a stationary point,
i.e., fulfilling the Karush-Kuhn-Tucker (KKT) optimality conditions
of the relaxed problem. Numerical experiments are carried out to evaluate
the proposed algorithm.

\section{System Model and Problem Formulation}

\subsection{System Models}

Consider a multiple-input single-output (MISO) downlink transmission
of C-RAN where several low-cost low-power RUs serve multiple single-antenna
users. Each RU is equipped with $M$ antennas. Let us denote by $B$
and $K$ the number of RUs and users in the network, respectively.
We assume that all RUs are connected to a common BBU through finite-capacity
fronthaul links. The BBU is also assumed to have all users' data and
(perfect) channel state information (CSI). We consider a fronthaul
network with compression strategy where BB signals for transmission
are precoded and compressed at the BBU before being forwarded to RUs.
Let $\{s_{k}\}_{k=1}^{K}$ be a set of the intended data to users
in which $s_{k}$ is a Gaussian input with unit energy, i.e., $\mathbb{E}[|s_{k}|^{2}]=1$.
Suppose that linear precoding is adopted at the BBU. The BB signal
generated for the transmission at RU $b$ is written as
\begin{equation}
{\bf x}_{b}=\sum_{k=1}^{K}{\bf w}_{b,k}s_{k}\label{eq:signal-at-CP}
\end{equation}
where ${\bf w}_{b,k}\in\mathbb{C}^{M\times1}$ is the beamformer from
RU $b$ to user $k$. ${\bf x}_{b}$ is then compressed and forwarded
to RU $b$ through the fronthaul link. We assume that the Gaussian
test channel is used to model the effect of compression on the fronthaul
link \cite{FronthaulCompressionCRAN-Mag}. Accordingly, the BB signal
at RU $b$ is given by
\begin{equation}
\hat{{\bf x}}_{b}={\bf x}_{b}+{\bf q}_{b}
\end{equation}
 where ${\bf q}_{b}\in\mathbb{C}^{M\times1}$ is the quantization
noise, which is independent of ${\bf x}_{b}$, and modeled as a complex
Gaussian distribution vector with covariance ${\bf Q}_{b,b}$, i.e.,
${\bf q}_{b}\sim\mathcal{CN}(0,{\bf Q}_{b,b})$. Note that ${\bf Q}_{b,b}$
is full-rank, i.e., $\text{rank}({\bf Q}_{b,b})=M$. We further assume
so called multivariate compression such that the compression of BB
signals for each RU is mutually dependent. Thus the associated quantization
noise vectors are correlated, i.e., ${\bf Q}_{b,i}=\mathbb{E}[{\bf q}_{b}{\bf q}_{i}^{H}]\neq0,\forall b\neq i$.
Based on the information theoretic formulation \cite[Ch. 9]{el2011network},
RU $b$ can successfully receive $\hat{{\bf x}}_{b}$ as long as the
following condition holds
\begin{equation}
\sum_{b=1}^{B}\log\left|{\bf Q}_{b,b}+\sum_{k=1}^{K}{\bf w}_{b,k}{\bf w}_{b,k}^{H}\right|-\log\left|{\bf Q}\right|\leq\sum_{b=1}^{B}\bar{C}_{b}\label{eq:BH:constraint}
\end{equation}
where
\[
{\bf Q}\triangleq\left[\begin{array}{cccc}
{\bf Q}_{1,1} & {\bf Q}_{1,2} & \cdots & {\bf Q}_{1,B}\\
{\bf Q}_{2,1} & {\bf Q}_{2,2} &  & {\bf Q}_{2,B}\\
\vdots &  & \ddots & \vdots\\
{\bf Q}_{B,1} & {\bf Q}_{B,2} & \cdots & {\bf Q}_{B,B}
\end{array}\right]
\]
is the compression covariance matrix and $\bar{C}_{b}$ is the capacity
of the frontlink between BBU and RU $b$. At RU $b,$ the BB signal
$\hat{{\bf x}}_{b}$ is transmitted to users through flat fading channels.
The received signal of user $k$ can be written as
\begin{align}
y_{k} & =\sum_{b=1}^{K}{\bf h}_{b,k}\hat{{\bf x}}_{b}+\sigma_{k}=\sum_{b=1}^{B}{\bf h}_{b,k}(\sum_{k=1}^{K}{\bf w}_{b,k}s_{k}+{\bf {\bf q}}_{b})+\sigma_{k}\nonumber \\
 & ={\bf h}_{k}{\bf w}_{k}+\sum_{j\neq k}^{K}{\bf h}_{k}{\bf w}_{j}+\sum_{b=1}^{B}{\bf h}_{b,k}{\bf q}_{b}+\sigma_{k}\label{eq:receivedsignal}
\end{align}
where ${\bf h}_{b,k}\in\mathbb{C}^{1\times M}$ is the (row) vector
representing the channel between RU $b$ and user $k$, and $\sigma_{k}\sim\mathcal{CN}(0,N_{0})$
is the additive white Gaussian noise at user $k$. In \eqref{eq:receivedsignal},
${\bf h}_{k}\triangleq[{\bf h}_{1,k},{\bf h}_{2,k},\ldots,{\bf h}_{B,k}]\in\mathbb{C}^{1\times MB}$
and ${\bf w}_{k}\triangleq[{\bf w}_{1,k}^{T},{\bf w}_{2,k}^{T},\ldots,{\bf w}_{B,k}^{T}]^{T}$
$\in\mathbb{C}^{MB\times1}$ denote the aggregate vectors of channels
and beamformers from all RUs to user $k$, respectively. We also denote
${\bf W}_{k}\triangleq{\bf w}_{k}{\bf w}_{k}^{H}\succeq0$, ${\bf W}_{k}\in\mathbb{C}^{MB\times MB}$,
$\text{rank}({\bf W}_{k})=1,\forall k$ and ${\bf w}_{b,k}{\bf w}_{b,k}^{H}={\bf T}_{b}{\bf W}_{k}{\bf T}_{b}^{T}$
where ${\bf T}_{b}\in\mathbb{R}_{+}^{M\times MB}$ is all-zero matrix
except the columns from $((b-1)M+1)$ to $(bM)$ which contain the
identity matrix. Suppose that single-user decoding is used and the
intercell-interference is treated as Gaussian noise. By the multivariate
compression strategy, the achievable rate for user $k$ is given by
\cite[Ch. 9]{el2011network}
\begin{equation}
\begin{alignedat}{1} & r_{k}({\bf W},{\bf Q})\triangleq\\
 & W\log\left(1+\frac{{\bf h}_{k}{\bf W}_{k}{\bf h}_{k}^{H}}{{\textstyle \sum_{j\neq k}^{K}}{\bf h}_{k}{\bf W}_{j}{\bf h}_{k}^{H}+{\bf h}_{k}{\bf Q}{\bf h}_{k}^{H}+WN_{0}}\right)
\end{alignedat}
\label{eq:rate}
\end{equation}
where $W$ is the bandwidth.

\subsection{RU-user Selection Scheme}

We can see from \eqref{eq:BH:constraint} and \eqref{eq:rate} that
there is a trade-off among power of the beamformers, quantization
noise covariances, and users' throughput under the finite-capacity
fronthaul links. More specifically, due to the constraint in (3),
it is not always possible to increase SE simply by using more transmit
power and/or making quantization noise variances small \cite{FronthaulCompressionCRAN-Mag}.
Therefore, we propose to employ an RU-user selection scheme where
each user is only served by neighboring RUs of significant strength.
The idea is that the power consumption are saved while diversity provided
by multicell cooperation is still exploited to increase the transmission
quality. In the selection scheme, beamformers between a BS and a user
is made to be zero if the corresponding link is not selected. Mathematically,
let us denote by $\phi_{b,k}\in\{0,1\}$ the selection preference
variable where $\phi_{b,k}=1$ indicates that RU $b$ serves user
$k$ and $\phi_{b,k}=0$ otherwise. The relation between beamformer
${\bf w}_{b,k}$ and variable $\phi_{b,k}$ is given by $\text{tr}({\bf T}_{b}{\bf W}_{k}{\bf T}_{b}^{T})\leq\phi_{b,k}u_{b,k}$
where $u_{b,k}$ is subject to a considered power constraint. Obviously,
$\phi_{b,k}=0$ implies ${\bf w}_{b,k}=0$.

\subsection{Power Consumption Model}

 Besides the data-dependent power consumption which is due to the
BB signal generation, i.e., $P_{\text{data}}\triangleq\frac{1}{\epsilon}(\sum_{b=1}^{B}\sum_{k=1}^{K}\text{tr}({\bf T}_{b}{\bf W}_{k}{\bf T}_{b}^{T})+\sum_{b=1}^{B}\text{tr}({\bf Q}_{b,b}))$
(where $\epsilon$ is power amplifier efficiency), we need to consider
all other sources of power which are spent for the network operation.
Those are generally referred as data-independent power consumption
which consists of the power consumed by operating the signal processing
circuits at the BBU, RUs, users and the fronthaul network, i.e.,
\begin{equation}
P_{0}\triangleq\sum_{b=1}^{B}\sum_{k=1}^{K}\phi_{b,k}P_{b,k}^{\text{cir}}+KP_{\text{Us}}+BP_{\text{RU}}\label{eq:power:consumption}
\end{equation}
where $P_{b,k}^{\text{cir}}$ is power for signal processing circuit
block of beamformer ${\bf w}_{b,k}$; $P_{\text{RU}}$ denotes the
circuit power consumed at an RU; and $P_{\text{Us}}$ is the circuit
power of a user.

\subsection{Problem Formulation}

We are interested in a  joint design of precoding, multivariate compression
and RU-user link selection that maximizes the network EE subject the
limited fronthaul capacity and per-RU power constraints, which is
formulated as\begin{subequations}\label{Prob:Gen:Problem}
\begin{align}
\underset{{\bf W},{\bf Q},\bm{\phi},{\bf u}}{\max}\ \  & f_{\text{EE}}\triangleq\frac{\sum_{k=1}^{K}r_{k}({\bf W},{\bf Q})}{P_{\text{data}}+P_{0}}\label{eq:obj:gen:prob}\\
\text{s.t. }\ \ \  & \text{tr}({\bf T}_{b}{\bf W}_{k}{\bf T}_{b}^{T})\leq\phi_{b,k}u_{b,k},\ \forall b,k\label{eq:soft power:precoder}\\
 & \sum_{k=1}^{K}u_{b,k}+\text{tr}({\bf Q}_{b,b})\leq\bar{P},\ \forall b\label{eq:power}\\
 & \sum_{b=1}^{B}\phi_{b,k}\geq1,\ \forall b,k\label{eq:min:connectivity}\\
 & \phi_{b,k}\in\{0,1\},\ \forall b,k,\label{eq:Boolean}\\
 & \begin{alignedat}{1} & \sum_{b=1}^{B}\log\left|{\bf Q}_{b,b}+\sum_{k=1}^{K}{\bf T}_{b}{\bf W}_{k}{\bf T}_{b}^{T}\right|\\
 & \qquad\qquad\qquad\qquad-\log\left|{\bf Q}\right|\leq\sum_{b=1}^{B}\bar{C}_{b}
\end{alignedat}
\label{eq:BH:constraint-1}\\
 & {\bf W}_{k}\succeq0,\ \forall k\label{eq:PSD}\\
 & \text{rank}({\bf W}_{k})=1,\ \text{rank}({\bf Q}_{b,b})=M,\ \forall b,k\label{eq:rank}
\end{align}
\end{subequations}where $\bm{\phi}\triangleq[\phi_{1,k},\ldots\phi_{b,k},\ldots,\phi_{B,K}]\in\{0,1\}^{BK}$.
Herein, \eqref{eq:power} is the power constraint with the power budget
$\bar{P}$ at RU $b$. \eqref{eq:min:connectivity} is added to ensure
that each user is always served by at least one RU. Clearly, problem
\eqref{Prob:Gen:Problem} is classified as rank-constrained mixed
Boolean nonconvex program for which a global optimum is challenging
to derive. Thus, a low-complexity solution is more preferable in practice.
Toward this end, we drop the rank constraint \eqref{eq:rank} and
base our proposed solution on the relaxed problem of \eqref{Prob:Gen:Problem}.
It is worth mentioning that problem \eqref{Prob:Gen:Problem} without
the rank constraints is still nonconvex and thus intractable.

\section{Proposed Solution}

We now propose an algorithm that finds a suboptimal solution of the
rank-relaxed problem of \eqref{Prob:Gen:Problem} based on the combination
of the SDP and DC programming, referred to as the SDP-DC algorithm.
In particular, the principle of the DC algorithm is used to iteratively
convexify the nonconvexity of the relaxed problem to achieve a sequence
of SDP formulations, whose solutions converge to a stationary point
of the rank-relaxed problem. To proceed, we note that the Boolean
constraint \eqref{eq:Boolean} can be equivalently rewritten as
\begin{equation}
\sum_{b=1}^{B}\sum_{k=1}^{K}(\phi_{b,k}^{2}-\phi_{b,k})\geq0,\ \phi_{b,k}\in[0,1].\label{eq:boolean:relax}
\end{equation}
It is easy to see that \eqref{eq:boolean:relax} actually implies
that $\phi_{b,k}\in\{0,1\}$. On the other hand, the objective of
\eqref{Prob:Gen:Problem} is a generic fractional function. To arrive
at a tractable formulation of the rank relaxed problem of \eqref{Prob:Gen:Problem},
we use the epigraph form to rewrite it as \cite{boyd2004convex}\begin{subequations}\label{Prob:relaxed:epi}
\begin{align}
\max_{\substack{{\bf W},{\bf Q},{\bf u},\bm{\phi},\\
\eta,t,{\bf z},{\bf g},{\bf q}
}
} & \quad\eta\\
\text{s. t.}\quad & t\eta\leq\sum_{k=1}^{K}z_{k}\label{eq:EE:epi}\\
 & z_{k}\leq\log(1+g_{k})\label{eq:rate:epi}\\
 & \begin{alignedat}{1}t\geq & \sum_{b=1}^{B}\sum_{k=1}^{K}(\frac{1}{\epsilon}u_{b,k}+\phi_{b,k}P_{b,k}^{\text{cir}})+KP_{\text{Us}}\\
 & \qquad\qquad\qquad+BP_{\text{RU}}+\frac{1}{\epsilon}\sum_{b=1}^{B}\text{tr}({\bf Q}_{b,b})
\end{alignedat}
\label{eq:power:consumption:epi}\\
 & g_{k}q_{k}\leq{\bf h}_{k}{\bf W}_{k}{\bf h}_{k}^{H}\label{eq:SINR:epi}\\
 & q_{k}\geq\sum_{j\neq k}^{K}{\bf h}_{k}{\bf W}_{j}{\bf h}_{k}^{H}+{\bf h}_{k}{\bf Q}{\bf h}_{k}^{H}+WN_{0}\label{eq:infererence:epi}\\
 & \eqref{eq:soft power:precoder},\eqref{eq:power},\eqref{eq:min:connectivity},\eqref{eq:BH:constraint-1},\eqref{eq:PSD},\eqref{eq:boolean:relax}
\end{align}
\end{subequations}where ${\bf z}=[z_{1},\ldots,z_{K}]$, ${\bf g}=[g_{1},\ldots,g_{K}]$
and ${\bf q}=[q_{1},\ldots,q_{K}]$. Further, in light of DC programming
(or concave-convex procedure), we rewrite \eqref{eq:soft power:precoder},
\eqref{eq:EE:epi} and \eqref{eq:SINR:epi} as
\begin{align}
(\phi_{b,k}-u_{b,k})^{2} & \leq(\phi_{b,k}+u_{b,k})^{2}-4\text{tr}({\bf T}_{b}{\bf W}_{k}{\bf T}_{b}^{T})\label{eq:soft-pow-precoder:rewrite}\\
(\eta+t)^{2} & \leq(\eta-t)^{2}+4\sum_{k=1}^{K}z_{k}\label{eq:EE:epi:rewrite}\\
(g_{k}+q_{k})^{2} & \leq(g_{k}-q_{k})^{2}+4{\bf h}_{k}{\bf W}_{k}{\bf h}_{k}^{H}\label{eq:SINR:epi:rewrite}
\end{align}
where the functions in both sides of the above constraints are convex,
which are amendable for the application of the DC algorithm. However,
direct applying DC algorithm to \eqref{Prob:relaxed:epi} always results
in an infeasible program. To understand this, let us recall constraint
\eqref{eq:boolean:relax} and replace the term $\phi_{b,k}^{2}$ by
its linear approximation at feasible point $\hat{\phi}_{b,k}\in\{0,1\}$
according to the DC algorithm, i.e.,
\begin{equation}
\Phi(\bm{\phi},\hat{\bm{\phi}})\triangleq\sum_{b=1}^{B}\sum_{k=1}^{K}(2\phi_{b,k}\hat{\phi}_{b,k}-(\hat{\phi}_{b,k})^{2}-\phi_{b,k})\geq0.\label{eq:slater}
\end{equation}
It is not difficult to check that the set $\{\phi_{b,k}\in[0,1]|\Phi(\bm{\phi},\tilde{\bm{\phi}})>0\}$
is empty. To cope with this issue, we apply a regularization technique
to arrive at the following program\begin{subequations}\label{Prob:relaxed:epi:relaxed}
\begin{align}
\max_{\substack{{\bf W},{\bf Q},{\bf u},\bm{\phi},\eta,t,{\bf z},{\bf g},{\bf q},\lambda}
} & \quad\eta-\alpha\lambda\\
\text{s. t.}\quad & \lambda+\sum_{b=1}^{B}\sum_{k=1}^{K}(\phi_{b,k}^{2}-\phi_{b,k})\geq0\label{eq:phi:epi:relax}\\
 & \eqref{eq:power},\eqref{eq:min:connectivity},\eqref{eq:BH:constraint-1},\eqref{eq:PSD},\eqref{eq:rate:epi},\eqref{eq:power:consumption:epi},\eqref{eq:infererence:epi},\\
 & \eqref{eq:soft-pow-precoder:rewrite},\eqref{eq:EE:epi:rewrite},\eqref{eq:SINR:epi:rewrite}
\end{align}
\end{subequations}by adding a slack variable $\lambda\geq0$ and
a penalty parameter $\alpha$. As can be seen, $\lambda$ allows \eqref{eq:slater}
to be satisfied for any $\hat{\phi}_{b,k}\in[0,1]$. In addition,
$\lambda$ is to be minimized in \eqref{Prob:relaxed:epi:relaxed}
and $\lambda=0$ immediately implies that an optimal solution of \eqref{Prob:relaxed:epi:relaxed}
is also feasible to \eqref{Prob:relaxed:epi}.

 We are now ready to propose a novel iterative algorithm that solves
\eqref{Prob:relaxed:epi:relaxed}. The central idea of the method
is to linearize the nonconvex parts of \eqref{eq:BH:constraint-1},
\eqref{eq:soft-pow-precoder:rewrite}\textendash \eqref{eq:SINR:epi:rewrite}
and \eqref{eq:phi:epi:relax} at each iteration to produce a sequence
of solutions that converge to a stationary point. Mathematical justification
of the proposed iterative approach is given in the following where
the superscript denotes the iteration index. We begin with the constraints
in \eqref{eq:soft-pow-precoder:rewrite}\textendash \eqref{eq:SINR:epi:rewrite}
which all have the same form $(x+y)^{2}+w$ of which a convex lower
bound is simply given by $2(x^{(n)}+y^{(n)})(x+y)-(x^{(n)}+y^{(n)})^{2}+w$
for any operating point $x^{(n)}$ and $y^{(n)}$. Thus, we can approximate
\eqref{eq:soft-pow-precoder:rewrite}\textendash \eqref{eq:SINR:epi:rewrite}
by the following second order cone constraints
\begin{align}
 & \begin{alignedat}{1} & (\phi_{b,k}-u_{b,k})^{2}\leq2(\phi_{b,k}^{(n)}+u_{b,k}^{(n)})(\phi_{b,k}+u_{b,k})\\
 & \qquad\qquad\qquad\quad-(\phi_{b,k}^{(n)}+u_{b,k}^{(n)})^{2}-4\text{tr}({\bf T}_{b}{\bf W}_{k}{\bf T}_{b}^{T})
\end{alignedat}
\label{eq:pow:precoderi:approx}\\
 & \begin{alignedat}{1} & (\eta+t)^{2}\leq2(\eta^{(n)}-t^{(n)})(\eta-t)\\
 & \qquad\qquad\qquad\quad-(\eta^{(n)}-t^{(n)})^{2}+4\sum_{k=1}^{K}z_{k}
\end{alignedat}
\label{eq:EE:epi:approx}\\
 & \begin{alignedat}{1} & (g_{k}+q_{k})^{2}\leq2(g_{k}^{(n)}-q_{k}^{(n)})(g_{k}-q_{k})\\
 & \qquad\qquad\qquad\quad-(g_{k}^{(n)}-q_{k}^{(n)})^{2}+4{\bf h}_{k}{\bf W}_{k}{\bf h}_{k}^{H}.
\end{alignedat}
\label{eq:SINR:epi:approx}
\end{align}
In the same manner, \eqref{eq:phi:epi:relax} can be replaced by
\begin{equation}
\lambda+\Phi(\bm{\phi},\bm{\phi}^{(n)})\geq0\label{eq:phi:relax:approx}
\end{equation}
Now we turn our attention to the remaining nonconvex constraint \eqref{eq:BH:constraint-1}
and denote $h_{b}({\bf W},{\bf Q})\triangleq\log|{\bf Q}_{b,b}+\sum_{k=1}^{K}{\bf T}_{b}{\bf W}_{k}{\bf T}_{b}^{T}|$.
Remark that $h_{b}({\bf W},{\bf Q})$ is jointly concave and differentiable
w.r.t. ${\bf W}$ and ${\bf Q}$ in domain $\{{\bf W},{\bf Q}\ |\ {\bf W}\succeq0,{\bf Q}\succeq0\}$.
This allows us for deriving the affine majorization of $h_{b}({\bf W},{\bf Q})$
\cite{DanTWCOM,JointprecodingMultivariate}, i.e.,
\[
\begin{alignedat}{1} & h_{b}^{(n)}({\bf W},{\bf Q};{\bf W}^{(n)},{\bf Q}^{(n)})\triangleq h_{b}({\bf W}^{(n)},{\bf Q}^{(n)})+\\
 & \sum_{k=1}^{K}\text{tr}({\bf T}_{b}^{T}({\bf T}_{b}(\sum_{k=1}^{K}{\bf W}_{k}^{(n)}){\bf T}_{b}^{T}+{\bf Q}_{b,b}^{(n)})^{-1}{\bf T}_{b}({\bf W}_{k}-{\bf W}_{k}^{(n)}))\\
 & +\text{tr}(({\bf T}_{b}({\textstyle \sum_{k=1}^{K}}{\bf W}_{k}^{(n)}){\bf T}_{b}^{T}+{\bf Q}_{b,b}^{(n)})^{-1}({\bf Q}_{b,b}-{\bf Q}_{b,b}^{(n)}))
\end{alignedat}
\]
which is the upper bound of $h_{b}({\bf W},{\bf Q})$, i.e., $h_{b}^{(n)}({\bf W},{\bf Q};{\bf W}^{(n)},{\bf Q}^{(n)})\geq h_{b}({\bf W},{\bf Q})$.
Again in the light of DC algorithm, \eqref{eq:BH:constraint-1} can
be replaced by the convex constraint
\begin{equation}
\sum_{b=1}^{B}h_{b}^{(n)}({\bf W},{\bf Q};{\bf W}^{(n)},{\bf Q}^{(n)})-\log|{\bf Q}|\leq\sum_{b=1}^{B}\bar{C}_{b}.\label{eq:BH:approx}
\end{equation}
Finally, problem \eqref{Prob:relaxed:epi:relaxed} at iteration $n+1$
of the proposed algorithm is approximated by the following convex
program
\begin{equation}
\begin{alignedat}{1}\max_{{\bf s}}\ \eta-\alpha^{(n)}\lambda\ \text{s.t.} & \{\eqref{eq:power},\eqref{eq:min:connectivity},\eqref{eq:PSD},\eqref{eq:rate:epi},\eqref{eq:power:consumption:epi},\eqref{eq:infererence:epi},\\
 & \ \ \ \ \eqref{eq:pow:precoderi:approx},\eqref{eq:EE:epi:approx},\eqref{eq:SINR:epi:approx},\eqref{eq:phi:relax:approx},\eqref{eq:BH:approx}\}
\end{alignedat}
\label{Prob:relaxed:epi:relaxed:approx}
\end{equation}
 where ${\bf s}\triangleq\{{\bf W},{\bf Q},{\bf u},\bm{\phi},\eta,t,{\bf z},{\bf g},{\bf q},\lambda\}$
denotes all the optimization variables. The proposed method is summarized
in Algorithm \ref{alg:proposedAlg}. In particular, the value of penalty
parameter $\alpha$ is not a constant in Algorithm \ref{alg:proposedAlg}
and the update of $\alpha$ (see step \ref{updatepenalty}) at each
iteration deserves some comments. In fact, $\alpha$ relates to the
degree of relaxation in \eqref{Prob:relaxed:epi:relaxed:approx},
i.e., a large $\alpha$ strongly forces $\lambda\rightarrow0$ leading
to $\phi_{b,k}\in\{0,1\}$, which implies more tightness for the selection
variable $\phi_{b,k}$ and vice versa. Thus, we initialize $\alpha^{(0)}$
by a small value to provide more searching space for $\bm{\phi}$,
and then gradually increase $\alpha$ with a factor $c>1$ until $\|\bm{\phi}^{(n+1)}-\bm{\phi}^{(n)}\|_{2}$
is small enough. This update rule provably ensures that $\alpha^{(n)}$
is bounded and $\underset{n\rightarrow+\infty}{\lim}\lambda^{(n)}=0$
as proved in Theorem \ref{thm:1}. Another observation is that if
Algorithm \ref{alg:proposedAlg} outputs ${\bf W}^{\ast}$ and ${\bf Q}^{\ast}$
satisfying $\text{rank}({\bf W}^{\ast})=1$ and $\text{rank}({\bf Q}_{b,b}^{\ast})=M$,
then ${\bf W}^{\ast}$ and ${\bf Q}^{\ast}$ are also feasible to
\eqref{Prob:Gen:Problem}. Since Algorithm \ref{alg:proposedAlg}
is derived on the rank-relaxed problem, it is highly likely that ${\bf Q}^{\ast}$
is full rank matrix. Also, we can prove that Algorithm \ref{alg:proposedAlg}
achieves the rank-1 solution of ${\bf W}_{k}^{\ast}$, but the detailed
proof is omitted due to the space limitation. The main idea of the
proof is briefly sketched as follows. We derive the dual problem of
\eqref{Prob:relaxed:epi:relaxed:approx} and show that the Lagrangian
multiplier corresponding to ${\bf W}_{k}\succeq0$ (denoted by ${\bf Z}_{k}$)
holds $\text{rank}({\bf Z}_{k})\geq MB-1$. Then by the KKT condition
${\bf Z}_{k}{\bf W}_{k}^{\ast}=0$, we arrive at $\text{rank}({\bf W}^{\ast})=1$.
The convergence of Algorithm 1 is studied in the following theorem.
\begin{thm}
\label{thm:1}There exists a finite positive integer $n_{0}$ such
that $\alpha^{(n)}=\alpha^{(n_{0})}$ for $n\geq n_{0}$, i.e., the
sequence $\{\alpha^{(n)}\}$ is bounded above and $\underset{n\rightarrow+\infty}{\lim}\lambda^{(n)}=0$.
In addition, Algorithm 1 generates a sequence of solutions converging
to a stationary point, i.e., fulfilling the KKT optimality conditions
of problem \eqref{Prob:relaxed:epi:relaxed:approx}.
\end{thm}
The proof of Theorem \ref{thm:1} is deferred to the Appendix. Since
$\underset{n\rightarrow+\infty}{\lim}\lambda^{(n)}=0$, the selection
variables converge to binary values eventually, and thus the solution
of Algorithm 1 also satisfies the KKT conditions of \eqref{Prob:relaxed:epi}.

\subsection*{Implementation Issues}

\begin{algorithm}[tb]
\caption{The proposed SDP-DC to solve \eqref{Prob:relaxed:epi:relaxed:approx}}
\label{alg:proposedAlg} \begin{algorithmic}[1]

\STATE \textbf{Initialization}: set $n:=0$, generate a set of initial
feasible value ${\bf s}^{(0)}$ of \eqref{Prob:relaxed:epi:relaxed:approx}
and initial penalty parameter $\alpha^{(0)}$.  \label{alg:initial}

\REPEAT

\STATE Solve \eqref{Prob:relaxed:epi:relaxed:approx} to obtain the
set of optimal values ${\bf s}^{\ast}$.  \label{alg:solveproblem}

\STATE Form the problem for the next iteration with ${\bf s}^{(n+1)}={\bf s}^{\ast}$

\STATE Update $\alpha^{(n+1)}:=c\alpha^{(n)}$ if $\|\bm{\phi}^{(n+1)}-\bm{\phi}^{(n)}\|_{2}>\varepsilon$.\label{updatepenalty}

\STATE$n:=n+1$

\UNTIL {Convergence }  \label{alg:end_iteration}

\STATE Obtain ${\bf W}^{\ast},{\bf Q}^{\ast},\bm{\phi}^{\ast},\eta^{\ast}$.

\end{algorithmic}
\end{algorithm}
We now discuss on some practical issues when implementing Algorithm
\ref{alg:proposedAlg}. As can be seen, convex program \eqref{Prob:relaxed:epi:relaxed:approx}
is classified as generic SDP due to the nonlinear constraints \eqref{eq:rate:epi}
and \eqref{eq:BH:approx}, and thus requires a high computational
complexity to solve. To obtain more computationally efficient formulations,
we can approximately convert \eqref{eq:rate:epi} and \eqref{eq:BH:approx}
into second order cone and linear matrix inequality constraints, and
thus the resulting programs are far more efficient to solve by modern
SDP solvers. More specifically, $\log|{\bf Q}|$ can be replaced by
a system of LMIs as in \cite[Sect. 4.18.d]{BenNemi:book:LectModConv}
and \cite[Lemma 1]{DoanhTCOM}, and \eqref{eq:rate:epi} can be approximated
by a system of conic quadratic constraints as in \cite{BenNemi:01:Onthepolyhedral},
\cite{Giang:15:JCOML}.
\begin{table}[tb]
\caption{Simulation Parameters }

\centering{}%
\begin{tabular}{c|c}
\hline
\noun{Parameters} & \noun{Value}\tabularnewline
\hline
\hline
Pathloss model & $37.6\log\left(d\text{ [km]}\right)+128.1$\tabularnewline
Log normal shadowing & 8 dB\tabularnewline
Cell radius & 750 m\tabularnewline
Number of RUs $B$ & 4\tabularnewline
Number of users & 8\tabularnewline
Number of Tx antennas $N$ & 2\tabularnewline
Signal bandwidth $W$ & 10 MHz\tabularnewline
Power amplifier efficiency $\epsilon$ & 0.35\tabularnewline
Power spectral density of noise $N_{0}$ & -174 dBm/Hz\tabularnewline
Circuit power for precoding $P_{b,k}^{\text{cir}}$ & 2 W\tabularnewline
Circuit power for an RU $P_{\text{RU}}$ & 17.5 dBW\tabularnewline
Circuit power for a user $P_{\text{Us}}$ & 20 dBm\tabularnewline
\hline
\end{tabular}\vspace{-2mm}\label{Tab. 1}
\end{table}

\section{Numerical results\label{Sect:Sim}}

We now provide the numerical experiments to demonstrate the effectiveness
of Algorithm \ref{alg:proposedAlg}. The general simulation parameters
are taken from \cite{JointNetworkBeamCoMPMixedConic} and listed in
Table \ref{Tab. 1}. The values of power budget $\bar{P}$ and fronthaul
capacity $\bar{C}_{b}$ are given in the caption of related figures.
To the best of our knowledge, the EE maximization (EEmax) problem
for this setting has not been investigated previously. For the comparison
purposes, we compare Algorithm \ref{alg:proposedAlg} with the one
in \cite[Alg. 1]{JointprecodingMultivariate}, which studies the SE
maximization (SEmax) for the same context.

\begin{figure}[tb]
\centering{}\includegraphics[width=0.8\columnwidth]{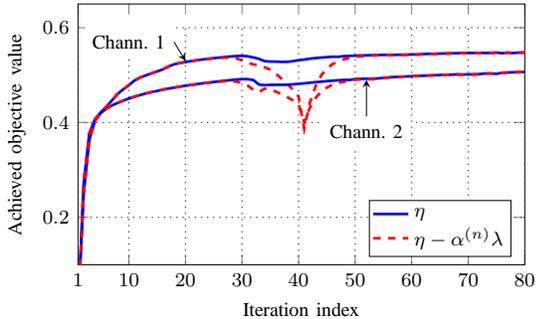}\vspace{-2mm}\caption{Convergence of Algorithm \ref{alg:proposedAlg} with $\bar{P}=40$dBm
and $\bar{C}_{b}=50$ Mnats/s. }
\vspace{-5mm}\label{fig.1}
\end{figure}

Fig. \ref{fig.1} shows the convergence behavior of Algorithm \ref{alg:proposedAlg}
for two random channel realizations by the objective of \eqref{Prob:relaxed:epi:relaxed:approx}
and the achieved EE. Remark that $\lambda^{(n)}>0$ for first iterations,
and thus $\alpha^{(n)}$ keeps increasing until reaching the limit.
As a result, the performance may be unstable at some intermediate
iterations due to the variation of the term $\alpha^{(n)}\lambda$.
After some point, $\alpha^{(n)}$ is fixed, and the latter iterations
lead to the stationary point.
\begin{figure}[tb]
\centering{}\includegraphics[width=0.8\columnwidth]{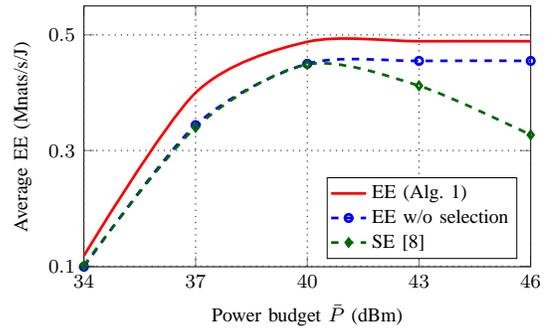}\vspace{-2mm}\caption{Average EE versus the per-RU transmit power budget $\bar{P}$ with
$\bar{C}_{b}=50$ Mnats/s. }
\vspace{-1mm}\label{fig.2}
\end{figure}
\begin{figure}[tb]
\centering{}\includegraphics[width=0.8\columnwidth]{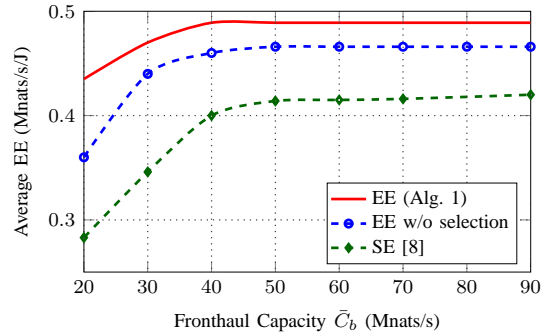}\vspace{-2mm}\caption{Average EE versus the fronthaul capacity $\bar{C}_{b}$ with $\bar{P}=43$
dBm. }
\vspace{-5mm}\label{fig.3}
\end{figure}

Fig. \ref{fig.2} compares the achieved EE versus the different transmit
power budgets for two strategies, i.e., maxSE in \cite{JointprecodingMultivariate}
and maxEE (Algorithm \ref{alg:proposedAlg}), both using multivariate
compression. We additionally illustrate the performance of EEmax without
using RU-user selection scheme to highlight the impact of the selection
strategy. Note that the numerical results on SEmax and EEmax comparison
presented in \cite{Dan:2013:SP:Fullduplex},\cite{DerrickKwanNg:2012:JWCOM:EE_OFDM}
imply that the EEmax (without selection) is in fact the SEmax in the
low-power regime, while in the high-power regime the SEmax reduces
and the EEmax remains unchanged. As can be seen, the performance shown
in Fig. \ref{fig.2} is consistent with those observations made previously.
On the other hand, Algorithm \ref{alg:proposedAlg} which adopts the
RU-user selection scheme outperforms the others for both low and high
power regions. This is easily understood since the selection mechanism
will switch off RU-user transmission links that do not offer a significant
improvement in achieved SE, saving power consumption remarkably.

Achieved EE versus the fronthaul capacity $\bar{C}_{b}$ is shown
in Fig. \ref{fig.3}. We can see that the achieved EE values for three
schemes increase following the increase of $\bar{C}_{b}$ due to the
fact that all BSs are allowed to transmit at higher data rates. However,
the EEs achieved by the EEmax strategies saturate after a certain
value of $\bar{C}_{b}$ (e.g., $\bar{C}_{b}=$40 Mnats/s for Algorithm
\ref{alg:proposedAlg}), while SEmax scheme keeps improving as $\bar{C}_{b}$
increases. In fact, SEmax scheme always uses all available power to
obtain more gain in achievable throughput. EEmax schemes, on the other
hand, aim at finding the optimal trade-off between achieved sum rate
and the total power consumption of the network.

\section{Conclusion}

We have considered a C-RAN downlink transmission where multivariate
compression fronthaul is adopted to generate the BB signals, which
are conveyed to RUs through limited capacity fronthaul links. We have
studied the joint design of precoding, multivariate compression and
RU-user selection that maximizes the EE measure. The optimization
problem is in fact a rank-constrained mixed Boolean nonconvex program.
We have applied relaxation techniques to drop the rank constraint
and convert the problem in a continuous domain. We have also used
DC programming to derive a low-complex iterative method to solve the
considered nonconvex continuous problem. The goal is to compute a
stationary point fulfilling the KKT optimality conditions. The effectiveness
of proposed algorithm has been demonstrated by the numerical results.

\section*{Appendix }

We prove the first claim by leveraging the result in \cite{le2014dc}.
For the ease of description, we pose problem \eqref{Prob:relaxed:epi:relaxed:approx}
at iteration $n+1$ in a general form as
\begin{equation}
{\textstyle \max_{{\bf s}\in{\cal S}^{(n)}({\bf s})}}\ \ f_{0}({\bf s})\ \ \text{s.t.}\ \{p({\bf s})\geq0\}\label{eq:prob:gen}
\end{equation}
where $f_{0}({\bf s})\triangleq\eta-\alpha^{(n)}\lambda$, $p({\bf s})\triangleq\lambda+\Phi(\bm{\phi},\bm{\phi}^{(n)})$
and ${\cal S}^{(n)}({\bf {\bf s}})$ is the feasible set, i.e., ${\cal S}^{(n)}({\bf s})\triangleq\{{\bf s}\ |\ \eqref{eq:power},\eqref{eq:min:connectivity},\eqref{eq:PSD},\eqref{eq:rate:epi},$
$\eqref{eq:power:consumption:epi},\eqref{eq:infererence:epi},\eqref{eq:pow:precoderi:approx},\eqref{eq:EE:epi:approx},\eqref{eq:SINR:epi:approx},\eqref{eq:BH:approx}\}$.
We also denote $I_{{\cal S}^{(n)}}({\bf s})$ and $\partial I_{{\cal S}^{(n)}}({\bf s})$
as the indicator function and the normal cone of ${\cal S}^{(n)}({\bf s})$
\cite{rockafellar2015convex}. Recall the KKT conditions for solution
of $\lambda$ at iteration $n+1$ which is given by
\begin{equation}
-\alpha^{(n)}+\mu^{(n+1)}+\vartheta^{(n+1)}=0;\ \lambda\vartheta^{(n+1)}=0;\ \lambda\geq0\label{eq:KKT:1}
\end{equation}
where $\mu^{(n+1)}$and $\vartheta^{(n+1)}$ are the Lagrangian multipliers
corresponding to $p({\bf s})\geq0$ and $\lambda\geq0$, respectively.
Let us assume $\underset{n\rightarrow+\infty}{\lim}\|\bm{\phi}^{(n+1)}-\bm{\phi}^{(n)}\|_{2}\neq0$.
One immediately has $\alpha^{(n)}\rightarrow+\infty$ and $\lambda>0$.
The latter is due to the fact that whenever $\lambda=0$, then $\phi_{b,k}\in\{0,1\},\forall b,k$
and $\|\bm{\phi}^{(n+1)}-\bm{\phi}^{(n)}\|_{2}=0$ (i.e., $\phi_{b,k}$
is fixed). Since $\lambda>0$ means $\vartheta^{(n+1)}=0$ by \eqref{eq:KKT:1},
then $\mu^{(n+1)}\rightarrow+\infty$. Next, we consider the KKT condition
for the Lagrangian function of \eqref{eq:prob:gen} given by
\begin{equation}
\nabla_{{\bf s}}f_{0}({\bf s}^{(n+1)})+\mu^{(n+1)}\nabla_{{\bf s}}p({\bf s}^{(n+1)})+\partial I_{{\cal S}^{(n)}}({\bf s}^{(n+1)})\in0\label{eq:KKT:2}
\end{equation}
If dividing \eqref{eq:KKT:2} by $\mu^{(n+1)}$ and let $n\rightarrow+\infty$,
one has $\nabla_{{\bf s}}p(\tilde{{\bf s}})+\partial I_{{\cal S}^{(n)}}(\tilde{{\bf s}})\in0$
which violates the Mangasarian\textendash Fromovitz constraint qualification,
i.e., $\nabla_{{\bf s}}p(\tilde{{\bf s}})$ and $\partial I_{{\cal S}^{(n)}}(\tilde{{\bf s}})$
are not positive-linearly independent at $\tilde{{\bf s}}$. This
implies the contradiction with assumption of $\alpha^{(n+1)}\rightarrow+\infty$,
and thus existing a finite integer $n_{0}$ such that $\alpha^{(n)}=\alpha^{(n_{0})}$
for $n\geq n_{0}$. Thereby we obtain $\|\bm{\phi}^{(n+1)}-\bm{\phi}^{(n)}\|_{2}=0$
leading to $\lambda=0$. Thus, $\underset{n\rightarrow+\infty}{\lim}\lambda^{(n)}=0$
which shows the first claim.

Next, we show that the limit point of Algorithm 1 fulfills the KKT
conditions. By the above arguments, $\phi_{b,k}^{(n)},\forall b,k$
converges into a Boolean set $\{0,1\}$ at iteration $n\geq n_{0}$.
At this point, follow exactly the convergence proof in \cite{MarksWright:78:AGenInnerApprox}
and remark that the feasible set of problem is bounded above by the
power constraint, we have the stationary point of Algorithm 1 guaranteeing
the KKT optimality conditions of \eqref{Prob:relaxed:epi:relaxed:approx}.
This completes the proof.\vspace{-2mm}

\section*{Acknowledgment}

This work was supported in part by the Academy of Finland under projects
Message and CSI Sharing for Cellular Interference Management with
Backhaul Constraints (MESIC) belonging to the WiFIUS program with
NSF, and Wireless Connectivity for Internet of Everything (WiConIE),
and the HPY Research Foundation. This project has been co-funded by
the Irish Government and the European Union under Ireland\textquoteright s
EU Structural and Investment Funds Programmes 2014-2020 through the
SFI Research Centres Programme under Grant 13/RC/2077.\vspace{-1mm}


\end{document}